\documentclass[aps,prb,twocolumn,amsmath,amssymb,floatfix]{revtex4-1}
\usepackage{tabularx}
\usepackage{bm}
\usepackage{euscript}
\usepackage{graphicx}
\usepackage{color}
\usepackage{amsfonts}
\usepackage{exscale}
\usepackage{amsbsy}
\usepackage{subfigure}
\usepackage{textcomp}

\begin{document}

\title{Entropy driven formation of a half-quantum vortex lattice}

\author{Suk Bum Chung and Steven A. Kivelson}
\affiliation{Department of Physics, Stanford University, Stanford, CA 94305}
\date{\today}

\begin{abstract}
Half-quantum vortices (HQVs) can exist in a superconductor or superfluid with an exact or approximate U(1)$\times$U(1) symmetry, for instance in spinor condensates, $^3$He-A, Sr$_2$RuO$_4$, and possibly cuprate superconductors with stripe order. In this paper, we show that a lattice of HQVs can be stabilized at finite temperature even when it does not have lower energy than the lattice of full vortices at $T=0$ since there is a gain in configurational entropy when a full vortex fractionalizes into a pair of HQVs. Specifically, the lattice of HQVs has an optical branch of phonon modes absent in the lattice of full vortices. Moreover, the HQV lattice at $T>0$ can have a different structure than  the HQV lattice at $T=0$.
\end{abstract}

\maketitle

\section{Introduction}

Superconductors and superfluids which require two phase variables to describe their condensates can be said to possess a U(1)$\times$ U(1) symmetry. One example, which has been known for a long time, is a thin film of the A-phase of $^3$He superfluid \cite{SALOMAA1985, BABAEV2005} in a perpendicular magnetic field, where the spin-triplet Cooper pairing allows for the possibility of the Cooper pair spin rotating in the plane perpendicular to the field.  More recently, Sr$_2$RuO$_4$ has been identified as an electronic analogue of $^3$He-A \cite{RICE1995,MACKENZIE2003} and as such can allow for a U(1)$\times$U(1) symmetry \cite{KEE2000, SARMA2006, CHUNG2007}. A U(1)$\times$U(1) symmetry can also arise in Bose condensates of atoms with nonzero integer spins, commonly known as `spinor condensates' \cite{STENGER1998, HO1998, OHMI1998, MUELLER2002, KASAMATSU2003, BARNETT2008}. Lastly, ``pair-density-wave'' (PDW) order - long known as the Fulde-Ferrell-Larkin-Ovchinnikov (FFLO) phase \cite{FULDE1964, Larkin1965} - can also allow for a U(1)$\times$U(1) symmetry as the superconducting and CDW phases are intertwined there. Recent, it was proposed \cite{BERG2007, BERG2009a, BERG2009} that experimental studies of the effects of the unidirectional spin and charge ordering (`stripe order') in the  cuprate superconductors  La$_{2-x}$Ba$_x$CuO$_4$ and  La$_{1.6-x}$Nd$_{0.4}$Sr$_x$CuO$_4$ \cite{LI2007, DING2008} shows evidence for the PDW order. Another candidate system for a PDW order is CeCoIn$_5$ under strong magnetic field \cite{RANDOVAN2003, BIANCHI2003}. Possible phase transitions in various systems with PDW order has been studied \cite{AGTERBERG2008, AGTERBERG2008a, BERG2009, RADZIHOVSKY2009}. 

A condensate with a U(1)$\times$U(1) symmetry allows for the existence of a half-quantum vortex (HQV) with only $\pi$, rather than $2\pi$, winding in the overall phase of the superfluid, $\theta$. For a charged superfluid, {\it i.e.} a superconductor,  a HQV binds flux of $h/4e$, which is half the flux of the conventional Abrisokov vortex, $\Phi_0 = h/2e$. This is a consequence of the existence of an extra phase variable $\alpha$, where at long distances, the condensate consists of two components with phases $(\theta \pm \alpha)/2$, regardless of the physics that gives rise to $\alpha$. The single-valuedness of the condensate wave function around a vortex is thus maintained even with $\Delta \theta = \pm \pi$ as long as it is also accompanied by a $\pm \pi$ winding in $\alpha$, thus making a HQV topologically stable. Conversely, observation of HQVs indicates the existence of the additional phase degree of freedom in the condensate, and thus serves as an indicator of the structure of the order parameter. The HQVs in chiral triplet superconductors have attracted particular interest in recent years stemming from the possibility that they will exhibit the simplest non-Abelian statistics \cite{READ2000, IVANOV2001, STERN2004, STONE2006}. More complex non-Abelian statistics of HQVs of spin-3/2 spinor condensate has also been studied \cite{Wu2009}.
 
\begin{figure}[bht]
\centerline{\includegraphics[width=.37\textwidth]{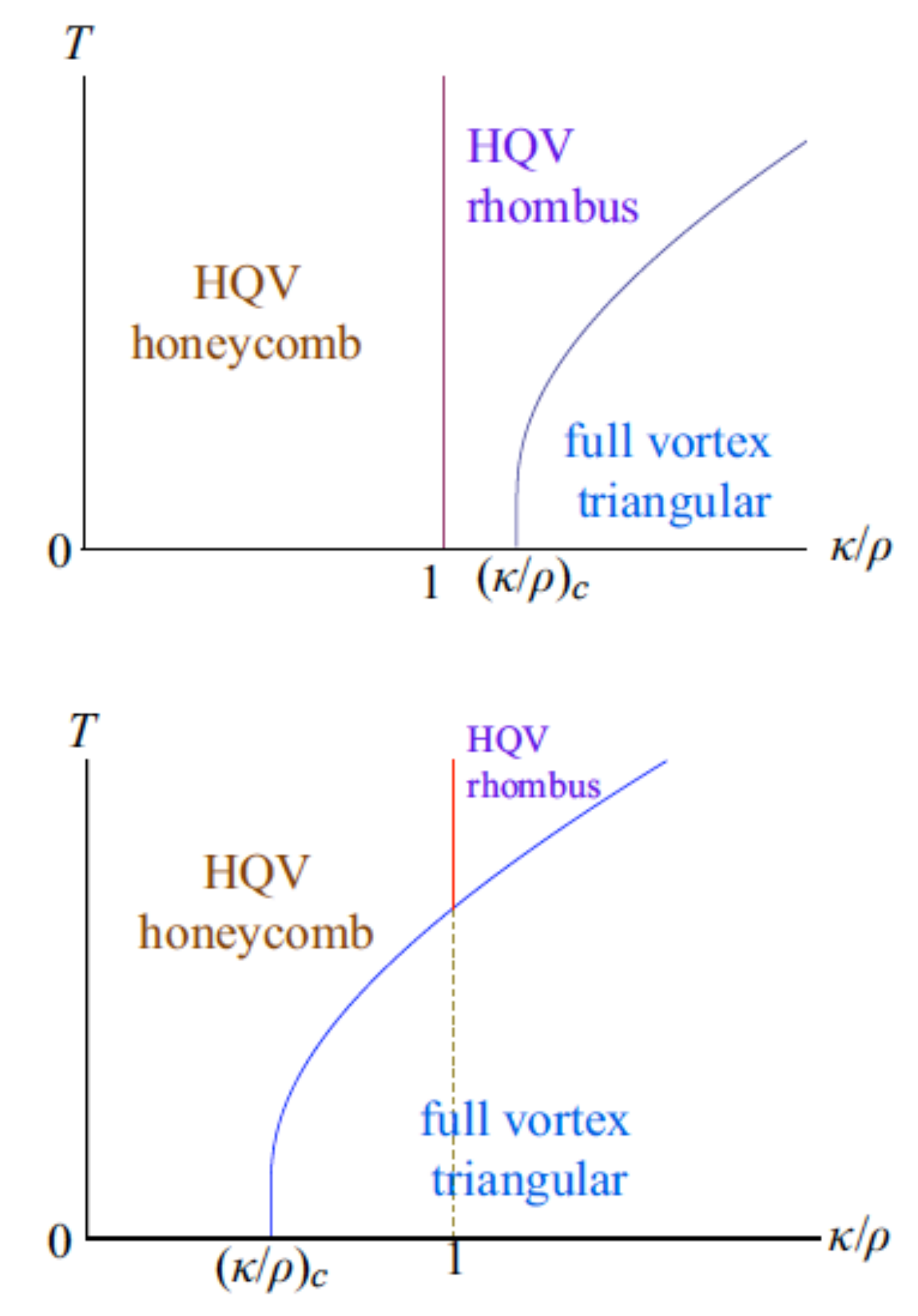}}
\caption{Two possible vortex lattice phase diagrams are possible, depending on the ratio of core energies; $T$ is temperature and $\kappa$ and $\rho$ are the phase stiffnesses for $\alpha$ and $\theta$, respectively. Note that without any change in the phase stiffness ratio $\kappa/\rho$, the HQV lattice can be stable at non-zero temperature even when it is not stable at $T=0$. $(\kappa/\rho)_c$ is increased if the core interaction between HQVs is repulsive, so the upper phase diagram is for the case that the vortex core interaction favors HQVs over full quantum vortices.}
\label{FIG:phase diagram}
\end{figure}

Although they are topologically stable, this does not guarantee that HQVs will ever occur in equilibrium systems.  Indeed, in superconductors, HQVs have never been detected in bulk, due to the energetics issues pointed out in Ref \onlinecite{CHUNG2007}. To the best of our knowledge, the recent apparent experimental detection of half-quantum fluxoids in mesoscopic Sr$_2$RuO$_4$ samples \cite{Jang2010} is, to date, the only successful experimental observation of a HQV in a single-crystal superconductor. 
However, in a lattice of HQVs, the lattice constant effectively plays the role of the system size, so a system close to $H_{c2}$ effectively becomes mesoscopic, as was shown in Ref. \onlinecite{CHUNG2009a}. The $T=0$ structure of the HQV lattice obtained for this case is  essentially the same as that obtained for spinor condensates in Refs. \onlinecite{MUELLER2002, BARNETT2008}. It is also analogous to the vortex--antivortex lattice configurations proposed in Refs. \onlinecite{GABAY1993,ZHANG1993} for a two-dimensional (2D) superfluid.

The main result of our paper is the vortex lattice phase diagram shown in Fig.~(\ref{FIG:phase diagram}) that shows the HQV lattice favored by non-zero temperature. Studies of the stability of the HQV lattice previously have focused on the $T=0$ case, {\it i.e.} on energy rather than free energy.  There has not been a thorough investigation of the possibility that the HQV lattice may be stabilized by its larger entropy. While Barnett, Mukerjee and Moore considered the possibility of a finite temperature transition from a full vortex lattice to a fractionalized vortex lattice \cite{BARNETT2008}, they only treated the case in which this is a result of the temperature dependence of the ratio of the stiffnesses of the two phase variables. Although this is ultimately also an entropic effect, it is special to the case in which $\alpha$ is associated with spin degrees of freedom with only weakly broken SU(2) invariance. In the present work, we will show that regardless of any physical origin of $\alpha$, entropy favors the HQV lattice, giving rise to a finite temperature phase transition from the full vortex lattice to the HQV lattice. In addition, we find the possibility that the entropy stabilized HQV lattice can have a different structure than the HQV lattice at $T=0$.  

The paper is organized as follows: In Section II, after comparing the free energy of a full vortex with a pair of HQVs, we show that a HQV lattice has higher entropy than a full vortex lattice by analyzing lattice phonon modes. In Section III, we show the energy competition between the triangular full vortex lattice and HQV lattices with different structures. In Section IV, we demonstrate that the HQV lattice at finite temperature is always dynamically stable in the thermodynamic limit. We conclude with a discussion in Section V.

\section{HQV entropy}


Results of this paper 
are based on the minimal model of a superconductor 
 in the London limit with a U(1)$\times$U(1) symmetry:
\begin{eqnarray}
\mathcal{H}[\theta,\alpha] 
=\frac{\hbar^2}{2M^*}\left[\rho\left(\nabla \theta - \frac{2e}{\hbar c}{\bf A}\right)^2 + \kappa(\nabla \alpha)^2\right] 
\label{EQ:HQV_Ham}
\end{eqnarray}
where $M^*$ is the Cooper pair mass, $\lambda = \frac{c}{4e}\sqrt{\frac{M^*}{\pi\rho}}$ is the London penetration depth, and $\rho$ and $\kappa$ are the stiffness for $\theta$ and $\alpha$, respectively. In this minimal model, we ignore any consideration 
of crystalline symmetry of the superconductor. We can treat the case of a neutral superfluid by setting $e \to 0$ (and consequently $\lambda \to \infty$) 
in 
all results.

\begin{figure}
\centerline{\includegraphics[width=.337\textwidth]{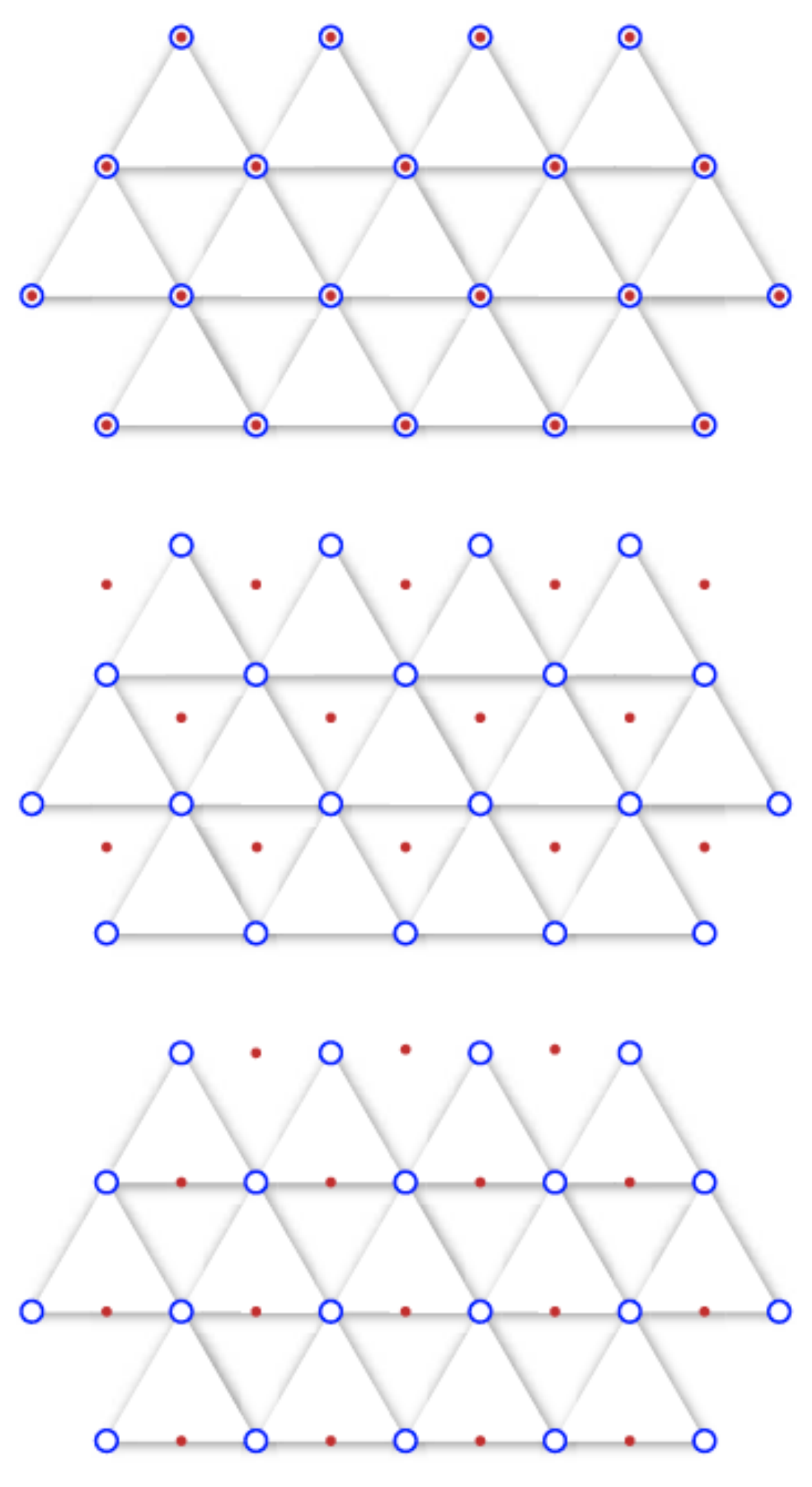}}
\caption{Three possible vortex lattice structures. The triangular lattice of full vortices is shown on the top. A full vortex can fractionalize into a pair of HQVs of opposite types, represented by a red dot inside a blue circle. Each type of HQV 
 forms a triangular lattice, but depending on the relative equilibrium displacements between the two types, 
the two different HQV lattice structures 
shown in the middle and the bottom are found.}
\label{FIG:vortex lattice}
\end{figure}

\subsection{Entropy argument}

We first show that a full quantum vortex with $(\Delta \theta, \Delta \alpha) = (2\pi, 0)$, that is the $2\pi$ phase winding only in the overall phase, can always lower its free energy by fractionalizing into a pair of HQVs, one with $(\Delta \theta, \Delta \alpha) = (\pi, \pi)$, the other with $(\Delta \theta, \Delta \alpha) = (\pi, -\pi)$, because this fractionalization will increase the configurational entropy. We emphasize here that this holds even in the case where the pair of HQVs have {\it higher} energy at $T=0$. As far as $T=0$ energy is concerned, 
in the limit $\lambda \to \infty$
\begin{eqnarray}
\epsilon^{full} &=& \pi \frac{\hbar^2}{M^*}\rho \ln \frac{R}{\xi},\nonumber\\
\epsilon^{HQV}_{pair}(r) &=& \frac{\pi}{2}\frac{\hbar^2}{M^*}\left[\kappa\ln\frac{r}{\xi}+\rho\left(\ln\frac{R}{\xi}-\ln\frac{r}{R}\right)\right],
\end{eqnarray}
where $R$ is the system size, 
 $r$ is the separation between two HQVs, and $\xi$ is the core radius, and we assume $R\gg r \gg \xi$ (in which limit the core energies are negligible).  
 The pair of HQVs have higher energy if $\kappa>\rho$. However, 
 dissociation of a full quantum vortex into a pair of HQVs results in 
 an entropy increase 
\begin{equation}
\Delta s (r) = \ln \frac{\pi r}{\xi},
\label{EQ:configEntropy}
\end{equation}
 from the number of the possible HQV pair configurations 
 with mean separation $r$. This implies that even when $\kappa > \rho$, 
 the free energy $f = \epsilon - Ts$ of the pair is lower at temperatures $T_{KT} \gg T>T_{dis} $ 
\begin{equation}
T_{dis} = \frac{\pi}{2}\frac{\hbar^2}{M^*}(\kappa - \rho) \sim T_{KT} \left(\frac{\kappa}{\rho}-1\right),
\end{equation}
where $T_{KT}$ is the Berezinskii-Kosterlitz-Thouless (BKT) temperature. 

This observation 
motivates us to compare the free energies of the full quantum vortex lattice and the HQV lattice in the case in which the latter has higher ground state energy. The regime we will examine is $H_{c1} \ll H \ll H_{c2}$, where screening of the charge current ({\it i.e.} the finite size of $\lambda$) is not 
 a dominant effect 
 but all vortex cores are still well separated. 

We begin with a heuristic argument:
It follows from the equipartition theorem that, in the limit $k_B T \gg \hbar \bar{\omega}$ 
where $\bar{\omega}$ is the Debye frequency for the phonons of the vortex lattice, the lattice entropy is
\begin{equation}
S 
=N k_B  \ln \frac{k_B T}{\hbar \bar{\omega}}
\end{equation}
where $N$ is the number of phonon modes.  
For fixed total number of $\theta$ vortices, $N_v$, piercing the system, the
HQV lattice has twice as many phonon modes, $N=2N_v$,  as the full quantum vortex 
lattice, since one full quantum vortex is topologically equivalent to a pair of HQVs. This means that the difference between the entropy of the HQV lattice and the full quantum vortex lattice is
\begin{eqnarray}
\Delta S &\approx& 2 N_v k_B \ln \frac{k_B T}{\hbar \bar{\omega}_{HQV}} - N_v k_B \ln \frac{k_B T}{\hbar \bar{\omega}_{full}}\nonumber\\
&=& N_v k_B\left[   \ln \frac{k_B T}{\hbar \bar{\omega}_{HQV}} - \ln \frac{\bar{\omega}_{HQV}}{\bar{\omega}_{full}}\right].
\label{EQ:latticeEntropy}
\end{eqnarray}
As we are in the classical regime, we can expect the first term to dominate over the second term, so $\Delta S >0$. Concerning the energy, we can take an estimate from the `single particle' picture we had in the last paragraph: 
\begin{equation}
\Delta E = N_v k_B T_{KT} \left[\frac{\kappa}{\rho}-\left(\frac{\kappa}{\rho}\right)_c\right],
\label{EQ:latticeEnergy0}
\end{equation} 
where $(\kappa/\rho)_c$ may be smaller or larger than 1 depending on the microscopic details ({\it i.e.} the ratio of the core energies of full and half vortices). From the thermodynamic expression $T_c \Delta S = \Delta E$ for a first order phase 
boundary 
and Eqs.\eqref{EQ:latticeEntropy} and \eqref{EQ:latticeEnergy0}, we 
find that the critical temperature for the transition from the full vortex lattice to the HQV lattice is
\begin{equation}
T_c \ln \frac{k_B T_c}{\hbar \bar{\omega}_{HQV}} = T_{KT}\left[\frac{\kappa}{\rho}-\left(\frac{\kappa}{\rho}\right)_c\right].
\label{EQ:Tc0}
\end{equation}
This equation indicates that for $0<\kappa/\rho - (\kappa/\rho)_c \ll 1$, we have $T_c \ll T_{KT}$. 

In the next subsection, we will calculate the free energy of the phonon modes of the full and HQV lattices and justify our argument in this subsection.

\subsection{Phonon modes of an HQV lattice}

While it is possible to imagine circumstances in which complicated vortex lattice structures occur, we will here make the simplifying assumption that 
the HQV lattice 
consists of two interpenetrating  triangular lattices, one of the $(\Delta \theta, \Delta \alpha) = (\pi, \pi)$ HQVs and the other of the  $(\Delta \theta, \Delta \alpha) = (\pi, -\pi)$ HQVs, translated by ${\bm \tau}$ with respect to each other. The full quantum vortex lattice in this picture 
is recovered in the limit  ${\bm \tau} \to {\bm 0}$. In this paper, we consider the cases with highest symmetry, as shown in Fig.(\ref{FIG:vortex lattice}): 
a honeycomb structure analogous to graphene or an interlaced rhombus where a $(\Delta \theta, \Delta \alpha) = (\pi, -\pi)$ HQV 
lies at a midpoint between two $(\Delta \theta, \Delta \alpha) = (\pi, \pi)$ HQVs. 
It is important to note here that, in an HQV lattice, there is a difference between `intra-lattice' interaction $U$ and the `inter-lattice' interaction $V$:
\begin{eqnarray}
U = \frac{\pi}{2}\frac{\hbar^2}{M^*}\rho \sum_i \sum_{j \neq i} &\frac{1}{2}\left[ K_0\left(\frac{r_{ij}}{\lambda}\right)- \frac{\kappa}{\rho}\ln\left(\frac{r_{ij}}{\xi}\right)\right],\nonumber\\
V = \frac{\pi}{2}\frac{\hbar^2}{M^*}\rho \sum_i \sum_j &\frac{1}{2}\left[ K_0\left(\frac{r_{ij'}}{\lambda}\right)+ \frac{\kappa}{\rho}\ln\left(\frac{r_{ij'}}{\xi}\right)\right],
\label{EQ:v-int}
\end{eqnarray}
where $i$ is the lattice point label for one of the two vortex lattices and ${\bf r}_{ij} = {\bf r}_i - {\bf r}_j$, ${\bf r}_{ij'} = {\bf r}_{ij} - \bm{\tau}$, ${\bf r}'_i = {\bf r}_i + \bm{\tau}$, and ${\bf r}_{i'j'} = {\bf r}'_i - {\bf r}'_j={\bf r}_{ij}$. Note that whereas $U$ is repulsive in both the $\theta$ channel and the $\alpha$ channel, $V$ is repulsive in the $\theta$ channel but attractive in the $\alpha$ channel.

The {\it vortex} equation of motion is 
\begin{equation}
{\bf F}_i \equiv  -\frac {\partial {\mathcal H}_v } {\partial {\bf r_i}}=\pi \hbar \rho {\bf \hat{z}} \times \frac{d}{dt}{\bf r}_i,
\label{EQ:Euler}
\end{equation}
which is equivalent to the Hamilton equations of motion with Hamiltonian 
\begin{equation}
{\mathcal H}_v(\{{\bf q}_i\})= U+V
\label{hamiltonian}
\end{equation}
where 
$q_{i}^x = \sqrt{\pi\hbar \rho} \ r_{i,x}$ and $q_{i}^y= \sqrt{\pi\hbar \rho}\ r_{i,y}$ are canonically conjugate variables.
One important consequence of this is that the number of phonon modes for a vortex lattice is only 
 half that of a usual crystalline system.

In order to 
obtain the phonon modes, we expand the equations of motion to 
first order in powers of the displacements, $\delta {\bf r}_i$, from the ground-state configuration. 
Because HQV lattice contains two vortices per unit cell, for each Bloch wave-vector  in the first Brillouin zone, 
there are two phonon branches, an acoustic and an optical.  
Near the Brillouin zone center, the acoustic modes involve 
 the motion of the HQV pair center of mass $\bar{\bf r}_i \equiv ({\bf r}_i + \bf r'_i)/2$ 
 while the optical modes  
 are associated with the relative motion of HQV, 
 $\Delta {\bf r}_i \equiv {\bf r}_i - {\bf r}'_i$.  (At large wave-vector, near the boundary of the Brillouin zone, these two motions are mixed, as shown in the Appendix.) The long wave-length acoustic modes can be obtained from the equation of motion
\begin{widetext}
\begin{eqnarray}
\pi \hbar \rho \frac{d}{dt}(\delta \bar{y}_i) &=& -\sum_{j \neq i}\left[\left(\frac{\partial^2 U}{\partial x_{ij}^2}+\frac{\partial^2 V}{\partial x_{ij'}^2}\right)\delta \bar{x}_{ij}+\left(\frac{\partial^2 U}{\partial x_{ij} \partial y_{ij}}+\frac{\partial^2 V}{\partial x_{ij'} \partial y_{ij'}}\right)\delta \bar{y}_{ij}\right],\nonumber\\
-\pi \hbar \rho \frac{d}{dt}(\delta \bar{x}_i) &=& -\sum_{j \neq i}\left[\left(\frac{\partial^2 U}{\partial y_{ij}^2}+\frac{\partial^2 V}{\partial y_{ij'}^2}\right)\delta \bar{y}_{ij}+\left(\frac{\partial^2 U}{\partial x_{ij} \partial y_{ij}}+\frac{\partial^2 V}{\partial x_{ij'} \partial y_{ij'}}\right)\delta \bar{x}_{ij}\right],
\label{EQ:EoM-ac}
\end{eqnarray}
\end{widetext}
and the corresponding optical modes from
\begin{widetext}
\begin{eqnarray}
\pi \hbar \rho \frac{d}{dt}(\delta\Delta y_i) &=& -\sum_{j \neq i}\left(\frac{\partial^2 U}{\partial x_{ij}^2}\delta\Delta x_{ij}+\frac{\partial^2 U}{\partial x_{ij} \partial y_{ij}}\delta\Delta y_{ij}\right) - \sum_j \left[\frac{\partial^2 V}{\partial x_{ij'}^2}(\delta\Delta x_i +\delta\Delta x_j)+\frac{\partial^2 V}{\partial x_{ij'} \partial y_{ij'}}(\delta\Delta y_i + \delta\Delta y_j)\right]\nonumber\\
&= &-\sum_{j \neq i}\left[\left(\frac{\partial^2 U}{\partial x_{ij}^2}-\frac{\partial^2 V}{\partial x_{ij'}^2}\right)\delta\Delta x_{ij}+\left(\frac{\partial^2 U}{\partial x_{ij} \partial y_{ij}}-\frac{\partial^2 V}{\partial x_{ij'} \partial y_{ij'}}\right)\delta\Delta y_{ij}\right]\nonumber\\
&-&2\sum_j\left[\frac{\partial^2 V}{\partial x_{ij'}^2}\delta\Delta x_i + \frac{\partial^2 V}{\partial x_{ij'} \partial y_{ij'}}\delta\Delta y_i\right],\nonumber\\
-\pi \hbar \rho \frac{d}{dt}(\delta\Delta x_i) &=& -\sum_{j \neq i}\left[\left(\frac{\partial^2 U}{\partial y_{ij}^2}-\frac{\partial^2 V}{\partial y_{ij'}^2}\right)\delta\Delta y_{ij}+\left(\frac{\partial^2 U}{\partial x_{ij} \partial y_{ij}}-\frac{\partial^2 V}{\partial x_{ij'} \partial y_{ij'}}\right)\delta\Delta x_{ij}\right]\nonumber\\
&-& 2\sum_j\left[\frac{\partial^2 V}{\partial y_{ij'}^2}\delta\Delta y_i + \frac{\partial^2 V}{\partial x_{ij'} \partial y_{ij'}}\delta\Delta x_i\right].
\label{EQ:EoM-opt}
\end{eqnarray}
\end{widetext}
This separation of relative and center of mass motion is justified in the limit where
\begin{eqnarray}
\frac{\partial^2 V}{\partial x_{ij'}^2} &\approx& \frac{\partial^2 V}{\partial x_{i'j}^2},\nonumber\\
\frac{\partial^2 V}{\partial y_{ij'}^2} &\approx& \frac{\partial^2 V}{\partial y_{i'j}^2},\nonumber\\
\frac{\partial^2 V}{\partial x_{ij'} \partial y_{ij'}} &\approx& \frac{\partial^2 V}{\partial x_{i'j} \partial y_{i'j}}.
\end{eqnarray}
Similar phonon dispersions were also found for the vortex-antivortex lattice in a 2D superfluid \cite{GABAY1993,ZHANG1993}.

Our result for dispersion of the acoustic branch justifies the arguments of Eq.\eqref{EQ:latticeEntropy}. We show in Appendix A that for the acoustic branch, we obtain the same dispersion relation that Tkachenko (for neutral superfluid $\lambda \to \infty$) \cite{Tkachenko1966} and Fetter \cite{FETTER1966, FETTER1967} obtained for the phonon modes of the full quantum vortex lattice:
\begin{equation}
\omega_a (q) \approx \frac{\pi \hbar n_v}{2M^*}\frac{\lambda d q^2}{\sqrt{1+(q\lambda)^2}},
\label{EQ:ac-dispersion}
\end{equation}
where $n_v = N_v/\Omega$ (where $\Omega$ is the total area of our vortex lattice) is the vortex density (where a HQV is counted as 1/2 a vortex) and $d$ is the distance between two closest HQVs of the same type. 
This derives from the fact that in the long-wavelength limit, the effective interaction for the acoustic branch is $U+V$. 
Because the acoustic dispersion is the same in both the full and HQV lattices, 
\begin{equation}
\Delta S = N_v k_B \ln \frac{k_B T}{\hbar \bar{\omega}_o},
\label{EQ:latticeEntropy1}
\end{equation}
where $\bar{\omega}_o$ is the 
mean frequency 
of the optical branch. 
Note the similarity of this expression to the heuristic result in Eq.\eqref{EQ:latticeEntropy}. 

The optical branch dispersion we obtain from Eq.\eqref{EQ:EoM-opt} justifies the modified version of Eq.\eqref{EQ:Tc0}
\begin{equation}
T_c \ln \frac{k_B T_c}{\hbar \bar{\omega}_o} = T_{KT}\left[\frac{\kappa}{\rho}-\left(\frac{\kappa}{\rho}\right)_c\right],
\label{EQ:Tc1}
\end{equation}
which we obtain from Eq.\eqref{EQ:latticeEntropy1}. This is because from the dispersion relation
\begin{equation}
\omega_o (q) \approx \frac{\pi \hbar n_v}{M^*}\sqrt{\left(1-\frac{\kappa^2}{\rho^2}\right)+\frac{\kappa^2}{4\rho^2}d^2 q^2},
\label{EQ:opt-dispersion}
\end{equation}
we obtain $k_B T_{KT}/\hbar \bar{\omega}_o \sim \rho/n_v \gg 1$,  which means the implied condition for Eq.\eqref{EQ:Tc1} - $k_B T_c \gg \hbar \bar{\omega}_o$ - is satisfied, unless $\kappa/\rho$ is fine-tuned to be very close to $(\kappa/\rho)_c$: $(\kappa/\rho) - (\kappa/\rho)_c \sim n_v/\rho$.  In order to avoid pathologies in the $T\to 0$ limit, we treat the phonons as a collection of quantum harmonic oscillators, within the the Debye approximation for their spectrum, from which we obtain  
\begin{equation}
T_c \ln \left[2 \sinh\frac{\hbar \bar{\omega}_o}{2k_B T_c}\right]^{-1} = \alpha T_{KT}\left[\frac{\kappa}{\rho}-\left(\frac{\kappa}{\rho}\right)_c\right],
\end{equation}
which was used to plot Fig.~(\ref{FIG:phase diagram}); $\alpha$ here is a $O(1)$ constant that can be obtained from the classical lattice energy calculation, which we will do in Section IV.

However, Eq.\eqref{EQ:opt-dispersion} also 
implies that at $T=0$ the HQV lattice is dynamically unstable  for any $\kappa > \rho$. 
In the asymptotic limit that the vortex lattice is very dilute, where $(\kappa/\rho)_c =1$, the $T=0$ transition between the two phases has the peculiar  feature that it is first order in the sense that there is a discontinuity in the first-derivative of the ground-state energy at the critical point, but there is also a soft mode with a frequency that vanishes at criticality and no regime in which the HQV lattice is metastable for $(\kappa/\rho)>(\kappa/\rho)_c$.  
If corrections due to the core energies make $(\kappa/\rho)_c < 1$, the $T=0$ transition is more typical of first order transitions, in that both the full and HQV lattices have a regime of metastability beyond the critical point.  
Below, we will show that thermal fluctuations can
modify the phonon dispersion and increase the range of dynamical stability.  Quantum fluctuations, which we have not considered seriously in the present paper, probably have a similar stabilizing effect, even at $T=0$, and thus eliminating these pathologies.

\section{Dynamic stability of lattice}

\subsection{``Self-consistent phonon'' method for $T>0$}

The dispersion relation obtained in the previous subsection is 
 modified 
 at finite temperature 
 by  the anharmonic part of the vortex interaction that we have so far ignored. In fact, we will show that when the anharmonic part is taken into consideration, the optical modes 
 of the HQV lattice are stabilized even for $\kappa > \rho$. 
 
To see this most simply, we treat the problem in a self-consistent phonon approximation.  Here we define a trial Hamiltonian which is quadratic in the deviations from the classical ground state configuration,
\begin{equation}
\mathcal{H}_{tr} = U_0 + \frac{1}{2!}\sum_{ij}\sum_{ab} K^{ab}_{ij} \delta q_i^a \delta q_j^b.
\end{equation}
where $U_0$ is the classical ground state energy, $ij$ refer to lattice sites and $ab$ are the coordinate components $x$ and $y$.  While at $T=0$, $K$ is equal to the appropriate matrix of second derivatives of ${\cal H}_v$, for $T>0$ we compute $K$ according the the Feynman-Jensen variational principle so as to minimize the variational free energy, $F'$:
\begin{equation}
F'[\{K^{ab}_{ij}\}]= F_{tr}+ \langle \mathcal{H} - \mathcal{H}_{tr} \rangle_{tr} \geq F
\end{equation}
where $F$ is the exact free energy, $\langle \ldots \rangle_{tr}$ is the thermal average in the Gibbs ensemble defined by $\mathcal{H}_{tr}$, and $F_{tr}$ is the free energy corresponding to $H_{tr}$~\cite{FEYNMAN1955}. With a little algebra, and with the use of Wick's theorem, it is easy to see that $K$ is determined self-consistently from the relation
\begin{eqnarray}
K^{ab}_{ij} &=& 2\frac{\partial \langle \mathcal{H} \rangle_{tr}}{\partial \langle \delta q^a_i \delta q^b_j \rangle_{tr}}
\label{EQ:self-consist}
\end{eqnarray}
Such approximation has been used to calculate phonon dispersion \cite{KUGLER1969,PLATZMAN1974}.  (We give a simple pedagogic example of this scheme applied to a single anharmonic oscillator in Appendix \ref{pedagogic}.)

To compute the first thermal corrections to the phonon dispersion, we expand $U[\{q_i^a\}]$ in powers of $\delta q_j^b$, keeping terms up through fourth order.  Then the self-consistency relation for the trial kernel can be expressed as:
\begin{eqnarray}
K^{ab}_{ij} 
= \frac{\partial^2 U}{\partial q_i^a \partial q_j^b}+\sum_{klcd}\frac{\langle \delta q_k^c \delta q_l^d \rangle_{tr}}2 \frac{\partial^4 U}{\partial q_k^c \partial q_l^d\partial q_i^a \partial q_j^b},\nonumber\\
\label{EQ:self-consist-quad}
\end{eqnarray}
where the derivatives of $U$ are evaluated at $\delta q^a_k=0$.

In general, at low temperatures, since $\langle \delta q_k^c \delta q_l^d \rangle_{tr}\propto T$, the thermal corrections to the phonon frequencies are small at low $T$.  The exception is as we approach a point of a zero temperature dynamical instability, where a mode goes soft, and the associated thermal fluctuations begin to diverge.  The key point about the variational approach (also illustrated by the simple pedagogic example in Appendix \ref{pedagogic}) is that the fluctuations are computed self-consistently, so that the finite $T$ regime of dynamical stability can even extend past the point of the $T=0$ instability.

\subsection{Scheme for self-consistent dispersion formula for the HQV lattice}

The calculation of phonon modes proceeds much as before.  Again, in the long wavelength limit, the acoustic mode is associated with the center-of-mass motion of the HQV pairs, while the optic mode is associated with the relative motion.  Following the same line of reasoning we followed in deriving Eq. \ref {EQ:EoM-ac} and \ref{EQ:EoM-opt}, the fluctuation operator can be expressed as a sum of contributions from the 
center-of-mass 
relative coordinate fluctuations in the long wavelength limit:
\begin{eqnarray}
\langle \delta q^a_{ij} \delta q^b_{ij} \rangle &=& \langle \delta q^a_{i'j'} \delta q^b_{i'j'} \rangle \nonumber\\
&=&\langle \delta \bar{q}^a_{ij} \delta \bar{q}^b_{ij} \rangle
 + \frac{\langle \delta \Delta q^a_i \delta \Delta q^b_i \rangle
 }{2}  -  \frac{\langle \delta \Delta q^a_j \delta \Delta q^b_i  \rangle
 }{2}  \nonumber\\
\langle \delta q^a_{ij'} \delta q^b_{ij'} \rangle &=& \langle \delta \bar{q}^a_{ij} \delta \bar{q}^b_{ij} \rangle
 + \frac{ \langle \delta \Delta q^a_i \delta \Delta q^b_i \rangle
 }{2} +  \frac{\langle \delta \Delta q^a_j \delta \Delta q^b_i  \rangle
}{2}.\nonumber
\label{EQ:AcOptDecomp}
\end{eqnarray}

The effect of these fluctuations on the long wave-length acoustic modes is minor. However, on the optic modes, the fluctuation induced stiffening of the effective spring-constants makes the HQV lattice  dynamically stable in the thermodynamic limit, at any finite temperature. Including the corrections to the effective stiffness from Eq.\eqref{EQ:self-consist-quad}, the frequency of optical phonon at $q=0$ becomes
\begin{equation}
\omega_o^2 (q=0) = \left(\frac{\pi \hbar n_v}{M^*}+\Delta \eta_o\right)^2 - \left(\frac{\kappa}{\rho}\frac{\pi \hbar n_v}{M^*}\right)^2,
\label{EQ:optGapT}
\end{equation}
where $\Delta \eta_o$ is a particular combination of fluctuation terms computed self-consistently. It is straightforward but complicated to compute this quantity, as we show explicitly in Appendix \ref{B}.  For instance, when the temperature is equal to the critical values at which the HQV vortex becomes thermodynamically stable,
\begin{equation}
\Delta \eta_o \sim \frac{\pi \hbar^2 \rho n_v}{8M^* \lambda^2}\langle \Delta x_i^2 + \Delta y_i^2\rangle.
\end{equation}
Generally, the effect of the thermal fluctuations is largest anywhere that the $T=0$ phonons become soft, as here the thermal fluctuations are largest.  Indeed, the impossibility of a dynamic instability at non-zero temperature can be proven by contradiction:  were there a putative point at which the optic mode became gapless, $\omega_o(q=0) \to 0$, the corresponding thermal fluctuations would diverge logarithmically with system size,  $\langle \Delta x_i^2 + \Delta y_i^2\rangle\sim T\ln N_v$.  

The same mechanism for attaining dynamic stability at non-zero temperature does apply in the neutral superfluid ({\it i.e.} in the limit $\lambda\to\infty$).  Possibly this signifies a dynamical instability to the formation of a HQV fluid, or it is possible that quantum effects or higher order non-linearities stabilize the HQV lattice.  We leave this issue to be resolved in future work.
In the next section, we will calculate the classical ground state energy for both full quantum vortex lattice and two HQV lattices of Fig.(\ref{FIG:vortex lattice}). This will allow us to 
explicitly derive Eq.\eqref{EQ:latticeEnergy0} and also determine what structure the HQV lattice should have.

\section{Vortex lattice structure}

We will first set up formulas for calculating the 
HQV lattice energy. We see from Eq.\eqref{EQ:v-int} that the classical lattice energy per unit area is
\begin{eqnarray}
E/\Omega &=& 
2n_v\left(\epsilon'_1 + U + V + U_0\right)\nonumber\\ 
&=& n_v\left[\epsilon_1+\delta\epsilon(\tau/\xi) \right]+ \frac{1}{2}n_v\frac{\pi\hbar^2}{M^*}\rho\left(\Sigma_0 + 
\frac{\kappa}{\rho}\Sigma_{sp}\right)\nonumber\\ 
\label{EQ:HQVfree}
\end{eqnarray}
for fixed magnetic field $B = n_v\Phi_0$, where $\epsilon'_1$, $\epsilon_1$ are the core energy of a  single HQV and a single full quantum vortex respectively, 
\begin{align}
\Sigma_0 =& \sum_{j \neq 0} K_0 \left( \frac{r_{0j}}{\lambda}\right) 
+ \sum_j K_0 \left( \frac{r_{0j'}}{\lambda}\right)\nonumber\\
\Sigma_{sp} =& \sum_j \ln\left(\frac{r_{0j'}}{\xi}\right) - \sum_{j \neq 0} \ln\left(\frac{r_{0j}}{\xi}\right)
\label{EQ:ESum}
\end{align} 
are dimensionless 
functions of  the lattice structure, and 
$\delta \epsilon(\tau/\xi)$ vanishes for $\tau/\xi \gg 1$, and  accounts for the difference between the core energies of the HQV and the full vortex when  $\tau/\xi \sim 1$. 
Note that $\Sigma_0$ 
and $\Sigma_{sp}$ have implicit dependence on $B$ 
since all distances between HQVs scale  as $r_{0j}, r_{0j'} \propto \sqrt{B}$.  

We find that the classical lattice Gibbs free energy, which determines the lattice structure, is proportional to the BKT temperature. 
This is because the quantity that can be tuned in a superconductor is the applied field
\begin{equation}
H = \frac {4\pi } \Omega\left(\frac{\partial E}{\partial B}\right).
\label{EQ:Ha}
\end{equation}
The thermodynamic relation between the interaction energy and the Gibbs free energy shows that the latter should be proportional to the BKT temperature:
\begin{eqnarray}
G/\Omega &=& E/\Omega - \frac{1}{4\pi}HB\nonumber\\ 
&=& -\frac{\pi\hbar^2 \rho}{4M^*} \frac{B}{\Phi_0} \left[\frac{1}{2} 
\Sigma_1 
 - \frac{\kappa}{\rho}\right]\nonumber\\
&=&  -n_v k_B T_{KT}\left[\frac{1}{2}
\Sigma_1 
- \frac{\kappa}{\rho}\right],
\label{EQ:lattice-energy}
\end{eqnarray}
where $B$ can be determined in terms of $H$ from Eq.\eqref{EQ:Ha} and 
\begin{align}
\Sigma_1 =& \sum_{j \neq 0}\left(\frac{r_{0j}}{\lambda}\right)^2 \left[K_2 \left( \frac{r_{0j}}{\lambda}\right) - K_0 \left( \frac{r_{0j}}{\lambda}\right)\right],\nonumber\\
+& \sum_{j}\left(\frac{r_{0j'}}{\lambda}\right)^2 \left[K_2 \left( \frac{r_{0j'}}{\lambda}\right) - K_0 \left( \frac{r_{0j'}}{\lambda}\right)\right]
\label{EQ:GSum}
\end{align}
is another dimensionless function of the lattice structure.

To determine the stability of different vortex lattice structures, we need to compute  the Gibbs free energy as a function of the applied field $H$. For a given applied field, $H$, depending on the vortex lattice structure, there will be a different total magnetic field $B$. We can see this by inverting Eq.\eqref{EQ:Ha}, assuming, as in Fig.~\ref{FIG:vortex lattice}, that the HQVs of the some type form triangular lattice. In terms of $H'_{c1} \equiv 8 \pi \epsilon'_1/\Phi_0$, 
ignoring terms that are higher order in $1/n_v\lambda^2$,
\begin{align}
B_{HQV} = H-H'_{c1} + & \frac{\Phi_0}{16\pi\lambda^2} \left[(1\!+\!\kappa/\rho)\ln\frac{4\pi(H-H'_{c1})\lambda^2}{\Phi_0}\right . \nonumber\\
&+(1\!-\!\kappa/\rho)\ln\frac{4\lambda^2}{\xi^2}\nonumber\\
&+(3\!+\!\kappa/\rho)-(3\!-\!\kappa/\rho)\gamma_E\nonumber\\
&\left.-(1\!-\!\kappa/\rho)\gamma_{FQV} -(1\!+\!\kappa/\rho)\gamma_{HQV}\right],\nonumber\\
\label{EQ:BfromH}
\end{align}
where, in terms of the exponential function ($E_1(x) \equiv \int^\infty_x dt e^{-t}/t$), $\gamma_{HQV}$ is a ${\bf \tau}$ dependent function defined as
\begin{align}
\gamma_{HQV} = &E_1(\pi n \tau^2) + \sum_{j\neq 0} E_1[\pi n({\bf r}_{0j} - {\bm \tau}^2)]\nonumber\\ 
+&\sum_{{\bf g}\neq 0} e^{i{\bf g}\cdot{\bm \tau}} \frac{e^{-{\bf g}^2/4\pi n}}{{\bf g}^2/4\pi n}, \label{EQ:struct}
\end{align}
$\gamma_E$ $(=0.5772\cdots)$ is 
Euler's constant, and
\begin{align}
\gamma_{FQV} = \sum_{(l,m)\neq(0,0)} &E_1[(2\pi/\sqrt{3})(l^2 + lm + m^2)]\nonumber\\ 
+ \sum_{(l,m)\neq(0,0)} &[(2\pi/\sqrt{3})(l^2 + lm + m^2)]^{-1}\nonumber\\
\times &\exp[(2\pi/\sqrt{3})(l^2 + lm + m^2)]\nonumber\\
\approx &0.07971.
\end{align}
From Eqs.\eqref{EQ:lattice-energy}  and \eqref{EQ:BfromH} ($\Sigma_1$ 
is evaluated in Appendix D), we obtain the Gibbs free energy 
\begin{widetext}
\begin{align}
\frac {G_{HQV}}{\Omega} 
&=-\frac{(H-H'_{c1})^2}{8\pi}-\frac{\Phi_0(H-H'_{c1})}{64\pi^2\lambda^2}\nonumber\\
&\times\left[\left(1+\frac{\kappa}{\rho}\right)\ln\frac{4\pi(H-H'_{c1})\lambda^2}{\Phi_0}+\left(1-\frac{\kappa}{\rho}\right)\ln\frac{4\lambda^2}{\xi^2}+2-\left(3-\frac{\kappa}{\rho}\right)\gamma_E-\left(1-\frac{\kappa}{\rho}\right)
\gamma_{HQV}-\left(1+\frac{\kappa}{\rho}\right)
\gamma_{FQV}\right],\nonumber\\
\label{EQ:HQV_latticeEnergy}
\end{align}
\end{widetext}
again ignoring terms that are higher order in $1/n_v\lambda^2$.

We now can derive the vortex lattice phase diagram shown in Fig.(\ref{FIG:phase diagram}) by minimizing this expression for the Gibbs free energy. Eq.\eqref{EQ:HQV_latticeEnergy} shows us that $\kappa/\rho$ determines energy competition between the HQV lattices with different structures, as $C_{HQV}$ is the only term in the equation that depends on the lattice structure. From Eq.\eqref{EQ:struct}, a straightforward numerical evaluation of the sums reveals that $\gamma_{HQV}$ smaller for the honeycomb structure than for the interlaced rhombus structure:  $\gamma_{HQV} \approx 0.4506$ for the former and $\gamma_{HQV} \approx 0.5379$ for the latter. Therefore, the honeycomb structure has  lower energy for $\kappa < \rho$ and the interlaced rhombus structure has lower energy for $\kappa > \rho$. Within our level of analysis, this tells us that if we obtain an HQV lattice at finite temperature through first order phase transition at $\kappa > \rho$, the lattice will have the interlace rhombus structure.

We can also determine the classical $(\kappa/\rho)_c$. For the triangular full vortex lattice, it is known \cite{FETTER1966a} that
\begin{align}
B_{FQV} = &H-H_{c1}\nonumber\\ 
+ &\frac{\Phi_0}{8\pi\lambda^2}\left[\log\frac{4\pi(H-H_{c1})\lambda^2}{\Phi_0}+2-\gamma_E-\gamma_{FQV}\right],\nonumber\\
\label{EQ:fullB}
\end{align}
and
\begin{align}
G_{FQV} / \Omega 
=&-\frac{(H-H_{c1})^2}{8\pi}\nonumber\\
&-\frac{\Phi_0(H-H_{c1})}{32\pi^2\lambda^2}\nonumber\\
&\times \left[\ln\frac{4\pi(H-H_{c1})\lambda^2}{\Phi_0}+1\!-\!\gamma_E\!-\!\gamma_{FQV}\right],\nonumber\\
\label{EQ:fullG}
\end{align}
where $H_{c1} \equiv 4\pi \epsilon_1/\Phi_0$. 
So the energy difference between the full vortex lattice and an HQV lattice is
\begin{eqnarray}
\Delta (G/\Omega) &\equiv& (G_{HQV} - G_{FQV})/\Omega\nonumber\\
&=&\frac{\Phi_0(H-H_{c1})}{64\pi^2\lambda^2}\left(\frac{\kappa}{\rho}-1\right)\nonumber\\
&\times&\left[\ln\frac{\Phi_0}{\pi(H-H_{c1})\xi^2}-\gamma_E-\gamma_{HQV}+\gamma_{FQV}\right]\nonumber\\ 
&+& \frac{(H-H_{c1})\Delta H_{c1}}{4\pi}\,
\end{eqnarray} 
where $\Delta H_{c1} = H'_{c1} - H_{c1}$. Note that since we are in the $H_{c1} \ll H \ll H_{c2}$ regime, $\Phi_0 \gg (H-H'_{c1})\xi^2$.  Thus,  if we ignore the core energy, we have $\Delta (G/\Omega)<0$ when $\rho > \kappa$ and $\Delta (G/\Omega)>0$ when $\rho < \kappa$.  If we include the core energy difference, the critical value will be
\begin{eqnarray}
&\,&1-(\kappa/\rho)_c\nonumber\\ 
&=& \frac{16\pi\lambda^2 \Delta H_{c1}}{\Phi_0}\left[\ln\frac{\Phi_0}{\pi(H-H_{c1})\xi^2}\!-\!\gamma_E\!-\!\gamma_{HQV}\!+\!\gamma_{FQV}\right]^{-1}.\nonumber\\
\label{EQ:criticalRatio}
\end{eqnarray}
We see from this equation that $(\kappa/\rho)_c$ approaches 1 at lower vortex density, {\it i.e.} as $H \to H_{c1}$. We also find, from Eqs\eqref{EQ:BfromH}, \eqref{EQ:fullB}, and \eqref{EQ:criticalRatio}, that there is a jump in $B$ for any finite temperature transition to the HQV lattice:
\begin{align}
\Delta B \equiv & B_{HQV} - B_{FQV}\nonumber\\
=& \frac{\Phi_0}{16\pi\lambda^2}\left[\frac{\kappa}{\rho}-\left(\frac{\kappa}{\rho}\right)_c\right]\nonumber\\
\times&\left[\ln\frac{\Phi_0}{\pi(H-H_{c1})\xi^2}-\gamma_E-\gamma_{HQV}+\gamma_{FQV}\right].
\end{align}


Presumably, there are also quantum corrections to $(\kappa/\rho)_c$ which we do not address here.

\section{Discussion}

We can classify methods for detecting this entropy driven formation of HQV lattice into direct and indirect. The direct method would be vortex imaging, either through measuring the local magnetic field distribution 
using a scanning SQUID or Hall magnetometer, or with 
neutron scattering. One signature from vortex imaging that should be examined is the vortex lattice structure, 
since, as we see in Fig.~(\ref{FIG:vortex lattice}), this is 
 different for a HQV lattice compared to the full quantum vortex lattice. 
 An indirect method would 
 be to look for evidence of a first order vortex lattice to vortex lattice transition as as a function of temperature. 


Throughout, we have assumed we have considered a system close to the zero temperature transition point, $|\kappa/\rho - (\kappa/\rho)_c| \ll 1$, where the transitions take place at temperatures small compared to $T_{KT}$.  At higher temperatures, other fluctuations will lead to a melting transition~\cite{FREY1994, BABAEV2004, SMORGRAV2005} of the vortex lattice itself.  When the above inequality is well satisfied, these two sorts of transition can be treated separately, as they occur on very different temperature scales.  Conversely, where it is not well satisfied, the physics of the full vortex lattice to HQV lattice transition may be altered significantly, and may even be preempted by melting of the full vortex lattice.

{\bf Acknowlegements: }  We owe special thanks to Sandy Fetter for patiently explaining to us his works on vortex lattice energetics and Sudip Chakravarty for teaching us how an unstable equilibrium can stabilize at finite temperature. We would also like to thank Egor Babaev, Ryan Barnett, Joel Moore, Subroto Mukerjee, Dror Orgad, Srinivas Raghu, Dan Agterberg, Eun-Ah Kim, Aharon Kapitulnik and Eytan Grosfeld for sharing their insights. We also thank the hosts and organizers of Gordon Research Conference 2010 on ``Strongly Correlated Electrons", Aspen Center for Physics Summer 2010 Program on ``Low Dimensional Topological Matter", and the Nordita program on ``Quantum Solids, Fluids and Gases"  where part of this work was completed. This work was supported in part by the DOE under contracts DE-AC02-76SF00515 (SBC) and DE-FG02-06ER46287 (SAK). 


\appendix
\section{Phonon dispersion}
To calculate normal modes from Eqs.\eqref{EQ:EoM-ac} and \eqref{EQ:EoM-opt}, we set ${\bf u}_i= (C_x, C_y) \exp[i({\bf q}\cdot{\bf r}_i - i\omega t)]$ where ${\bf u} =\bar{{\bf r}_i}$ or $\delta\Delta {\bf r}_i $. For both acoustic and optical modes, we can write down the normal mode equation of motion in the form
\begin{equation}
-i\omega \left(\begin{matrix}C_y \\ -C_x\end{matrix}\right) = -\left(\begin{array}{cc}\eta({\bf q})+\zeta({\bf q}) & \alpha({\bf q})\\
\alpha({\bf q}) & \eta({\bf q})-\zeta({\bf q}) \end{array}\right)\left(\begin{matrix}C_x \\ C_y\end{matrix}\right),
\end{equation}
which gives us the frequency
\begin{equation}
\omega^2 ({\bf q}) = \eta^2 ({\bf q}) - \zeta^2 ({\bf q}) - \alpha^2 ({\bf q}).
\label{EQ:dispersionPara}
\end{equation}
As for these quantities $\eta, \zeta$ and $\alpha$, we obtain for acoustic modes
\begin{eqnarray}
\eta_a({\bf q}) &=& \frac{1}{2\pi \hbar \rho}\sum_{j \neq i}[(\mathcal{T}_{ij}U) + (\mathcal{T}_{ij'}V)](1-e^{-i{\bf q}\cdot{\bf r}_{ij}}),\nonumber\\
\zeta_a({\bf q}) &=& \frac{1}{2\pi \hbar \rho}\sum_{j \neq i}[(\mathcal{\tilde{T}}_{ij}U)\cos 2\phi_{ij} + (\mathcal{\tilde{T}}_{ij'}V)\cos 2\phi_{ij'}]\nonumber\\
&\times&(1-e^{-i{\bf q}\cdot{\bf r}_{ij}}),\nonumber\\
\alpha_a({\bf q}) &=& \frac{1}{2\pi \hbar \rho}\sum_{j \neq i}[(\mathcal{\tilde{T}}_{ij}U)\sin 2\phi_{ij} + (\mathcal{\tilde{T}}_{ij'}V)\sin 2\phi_{ij'}]\nonumber\\
&\times&(1-e^{-i{\bf q}\cdot{\bf r}_{ij}}),
\label{EQ:acPara0}
\end{eqnarray}
and for optical modes
\begin{eqnarray}
\eta_o({\bf q}) &=& \frac{1}{2\pi \hbar \rho}\sum_{j \neq i}[(\mathcal{T}_{ij}U )- (\mathcal{T}_{ij'}V)](1-e^{-i{\bf q}\cdot{\bf r}_{ij}})\nonumber\\ 
&+& \frac{1}{\pi \hbar \rho}\sum_j \mathcal{T}_{ij'}V,\nonumber\\
\zeta_o({\bf q}) &=& \frac{1}{2\pi \hbar \rho}\sum_{j \neq i}[(\mathcal{\tilde{T}}_{ij}U)\cos 2\phi_{ij} - (\mathcal{\tilde{T}}_{ij'}V)\cos 2\phi_{ij'}]\nonumber\\
&\times&(1-e^{-i{\bf q}\cdot{\bf r}_{ij}})\nonumber\\ 
&+& \frac{1}{\pi \hbar \rho}\sum_j  (\mathcal{\tilde{T}}_{ij'}V)\cos 2\phi_{ij'},\nonumber\\
\alpha_o({\bf q}) &=& \frac{1}{2\pi \hbar \rho}\sum_{j \neq i}[(\mathcal{\tilde{T}}_{ij}U)\sin 2\phi_{ij} - (\mathcal{\tilde{T}}_{ij'}V)\sin 2\phi_{ij'}]\nonumber\\
&\times&(1-e^{-i{\bf q}\cdot{\bf r}_{ij}})\nonumber\\ 
&+& \frac{1}{\pi \hbar \rho}\sum_j  (\mathcal{\tilde{T}}_{ij'}V)\sin 2\phi_{ij'},
\label{EQ:optPara0}
\end{eqnarray}
where
\begin{align}
\mathcal{T} &= \frac{\partial^2}{\partial x^2} + \frac{\partial^2}{\partial y^2} =  \frac{1}{r}\frac{d}{dr}\left(r \frac{d}{dr}\right),\nonumber\\
\tilde{\mathcal{T}} &= \left(\frac{\partial^2}{\partial x^2}\!-\!\frac{\partial^2}{\partial y^2}\right)\cos 2\phi + 2\frac{\partial^2}{\partial x \partial y} \sin 2\phi = r\frac{d}{dr}\left(\frac{1}{r} \frac{d}{dr}\right),
\end{align}
and $\tan \phi = y/x$. 

In the continuum limit, for the acoustic phonon parameters we obtain
\begin{eqnarray}
\eta_a(q) &=& \frac{\hbar n_v}{2 M^*}\int r dr \int d\phi \mathcal{T} \left[K_0\left(\frac{r}{\lambda}\right)\right]\nonumber\\
&\times&(1-e^{-iqr\cos\phi})\nonumber\\ 
&=& \frac{\pi\hbar n_v}{M^* \lambda^2}\int r dr [1-J_0(qr)] K_0\left(\frac{r}{\lambda}\right)\nonumber\\ 
&\approx& \frac{\pi \hbar n_v}{M^*}\frac{(q\lambda)^2}{1+(q\lambda)^2}
\label{EQ:acPara011}
\end{eqnarray}
and
\begin{eqnarray}
|\zeta_a(q) + i \alpha_a(q)| &=&  \frac{\hbar n_v}{2 M^*}\int r dr \int  d\phi  e^{i2\phi}\mathcal{\tilde{T}} \left[K_0\left(\frac{r}{\lambda}\right)\right]\nonumber\\ 
&\times& (1-e^{-iqr\cos\phi})\nonumber\\
&=& \frac{\pi \hbar n_v}{M^* \lambda^2} \int^\infty_d r dr J_2 (qr) K_2\left(\frac{r}{\lambda}\right)\nonumber\\ 
&\approx& \frac{\pi \hbar n_v}{M^*}\left[\frac{(q\lambda)^2}{1+(q\lambda)^2}-\frac{1}{8}q^2 d^2\right].
\label{EQ:acPara012}
\end{eqnarray}
and for the optical phonon parameter
\begin{widetext}
\begin{eqnarray}
\eta_o(q) &=& \frac{\hbar n_v}{2 M^*}\int r dr d\phi (1-e^{-iqr\cos\phi}) \mathcal{T} \left[-\frac{\kappa}{\rho}\log\left(\frac{r}{\xi}\right)\right]  + \frac{\hbar n_v}{2 M^*}\int r dr d\phi \mathcal{T}\left[K_0\left(\frac{r}{\lambda}\right)+\frac{\kappa}{\rho}\log\left(\frac{r}{\xi}\right)\right]\nonumber\\
&=& \frac{\hbar n_v}{2M^* \lambda^2}\int r dr \int d\phi K_0\left(\frac{r}{\lambda}\right) \approx \frac{\pi \hbar n_v}{M^*}
\label{EQ:optPara011}
\end{eqnarray}
and
\begin{eqnarray}
|\zeta_o(q) + i \alpha_o(q)| &=& \frac{\hbar n_v}{2 M^*}\int r dr d\phi e^{i2\phi}(1-e^{-iqr\cos\phi})  \mathcal{\tilde{T}} \left[-\frac{\kappa}{\rho}\log\left(\frac{r}{\xi}\right)\right]  + \frac{\hbar n_v}{2 M^*}\int r dr d\phi e^{i2\phi} \mathcal{\tilde{T}}\left[K_0\left(\frac{r}{\lambda}\right)+\frac{\kappa}{\rho}\log\left(\frac{r}{\xi}\right)\right]\nonumber\\
&=& \frac{\hbar n_v}{M^*}\frac{\kappa}{\rho}\int^\infty_d dr \frac{J_2 (qr)}{r} \approx \frac{\pi \hbar n_v}{M^*}\frac{\kappa}{\rho}\left(1-\frac{1}{8}q^2 d^2\right).
\label{EQ:optPara012}
\end{eqnarray}
\end{widetext}
We obtain Eqs.\eqref{EQ:ac-dispersion} and \eqref{EQ:opt-dispersion} by inserting Eqs.\eqref{EQ:acPara011}, \eqref{EQ:acPara012}, and Eqs.\eqref{EQ:optPara011}, \eqref{EQ:optPara012}, respectively, into Eq.\eqref{EQ:dispersionPara}. To obtain finite temperature modification of dispersion, we therefore need to calculate the finite temperature correction to these phonon parameters.

\section{Pedagogic example of the self-consistent phonon approximation}
\label{pedagogic}
We can show from an explicit example that at high enough temperature, the self-consistent harmonic approximation around an {\it unstable} equilibrium may be more accurate than the approximation taken around a stable equilibrium. Our example is a particle in 
a two-well quartic potential,
\begin{equation}
\mathcal{H} = \frac{1}{2m}p^2 - \frac{1}{2}kq^2 + \frac{1}{4}\lambda q^4,
\end{equation}
with $k,\lambda>0$. At finite temperature, we can either take a self-consistent harmonic approximation around $q=0$ (which is the point of stable equilibrium for $k <0$ and of unstable equilibrium for $k >0$)
\begin{equation}
\mathcal{H}_{tr1} = \frac{1}{2m}p^2 + \frac{1}{2}Kq^2,
\end{equation}
where self-consistently $K = -k/2 + \sqrt{k^2/4 + 3\lambda k_B T/2}$, or (for $k >0$) around a point of   stable equilibrium
\begin{equation}
\mathcal{H}_{tr2} = \frac{1}{2m}p^2 + \frac{1}{2}K'(q-q_0)^2,
\end{equation}
where self-consistently $q_0 = \pm \sqrt{k/\lambda}$ and $K' = k + \sqrt{k^2 + 3\lambda k_B T/2}$. The Feynman-Jensen free energy in these two cases are 
\begin{eqnarray}
F'_1 &=& -k_B T \ln Z_{tr1} + \langle \mathcal{H} - \mathcal{H}_{tr1}\rangle_{tr1}\nonumber\\
&=& - k_B T \ln 2\pi k_B T \sqrt{\frac{m}{K}},\nonumber\\
F'_2 &=& -k_B T \ln Z_{tr2} + \langle \mathcal{H} - \mathcal{H}_{tr2}\rangle_{tr2}\nonumber\\
&=& - k_B T \ln 2\pi k_B T \sqrt{\frac{m}{K'}}-k_B T \frac{k^2}{4\lambda},
\end{eqnarray}
respectively. 

Note that both solutions are at least metastable for  all $T>0$.  For $T>T'$, the harmonic approximation around $q=0$ has lower variational free energy than the one around $q=q_0$, where 
$T'$ is the solution of the implicit equation,
\begin{equation}
\frac{k^2}{2\lambda k_B T'} = \ln \frac{K'}{K} = \ln \frac{2\sqrt{1+3\lambda k_B T'/2k^2}+2}{\sqrt{1 + 6\lambda k_B T'/k^2}-1}.
\end{equation}

\section{Optical phonon stabilization at $T>0$}
\label{B}

Here we examine the gap of the optical phonon (that is, the optical phonon frequency at $q=0$) at finite temperature. We can expect the optical phonon stabilization to occur at the temperature where the optical phonon is gapless. Note that translational invariance ensures that acoustic phonon remains gapless at finite temperature. In order to calculate this, as indicated by Eq.\eqref{EQ:self-consist-quad}, we apply the operator 
 \begin{eqnarray}
&\,&1+\frac{1}{2}\langle \delta x^2\rangle \frac{\partial^2}{\partial x^2} + \langle \delta x \delta y \rangle \frac{\partial^2}{\partial x \partial y} + \frac{1}{2}\langle \delta y^2\rangle \frac{\partial^2}{\partial y^2}\nonumber\\ 
&=& 1+\frac{1}{4}(\langle \delta x^2\rangle + \langle \delta y^2\rangle)\left(\frac{\partial^2}{\partial x^2}+\frac{\partial^2}{\partial y^2}\right)\nonumber\\ 
&+& \frac{1}{4}(\langle \delta x^2\rangle - \langle \delta y^2\rangle)\left(\frac{\partial^2}{\partial x^2}-\frac{\partial^2}{\partial y^2}\right) + \langle \delta x \delta y \rangle \frac{\partial^2}{\partial x \partial y}\nonumber\\
\end{eqnarray}
on the vortex-vortex interactions $U$ and $V$ in Eq.\eqref{EQ:optPara0}. 
We find
\begin{widetext}
\begin{eqnarray}
2\pi \hbar \rho\eta_o(q=0) &=& \frac{2\pi^2 \hbar^2 \rho n_v}{M^*}\nonumber\\
&+& \frac{1}{2}\sum_j [\langle \delta x_{ij'}^2 + \delta y_{ij'}^2\rangle\mathcal{T}_{ij'}+(\langle \delta x_{ij'}^2 - \delta y_{ij'}^2\rangle \cos 2\phi_{ij'}+ 2\langle \delta x_{ij'}\delta y_{ij'}\rangle \sin 2\phi_{ij'})\tilde{\mathcal{T}}_{ij'}]\mathcal{T}_{ij'}V,\nonumber\\
&=& \frac{2\pi^2 \hbar^2 \rho n_v}{M^*}\nonumber\\
&+& \frac{1}{4}\langle \delta \Delta x_i^2 + \delta \Delta y_i^2\rangle \sum_j\mathcal{T}_{ij'}^2 V + \frac{1}{4} \sum_j \langle \delta \Delta x_i \delta \Delta x_j + \delta \Delta y_i \delta \Delta y_j\rangle\mathcal{T}_{ij'}^2 V\nonumber\\
&+&\frac{1}{4}\sum_j (\langle \delta \Delta x_i \delta \Delta x_j - \delta \Delta y_i \delta \Delta y_j\rangle \cos 2\phi_{ij'}+ 2\langle \delta \Delta x_i\delta \Delta y_j\rangle \sin 2\phi_{ij'})\tilde{\mathcal{T}}_{ij'}\mathcal{T}_{ij'} V\nonumber\\
&+& \frac{1}{2}\sum_j [\langle \delta \bar{x}_{ij}^2 + \delta \bar{y}_{ij}^2\rangle\mathcal{T}_{ij'}+(\langle \delta \bar{x}_{ij}^2 - \delta \bar{y}_{ij}^2\rangle \cos 2\phi_{ij'}+ 2\langle \delta \bar{x}_{ij'}\delta \bar{y}_{ij'}\rangle \sin 2\phi_{ij'})\tilde{\mathcal{T}}_{ij'}]\mathcal{T}_{ij'}V,\nonumber\\
2\pi \hbar \rho |(\zeta_o + i\alpha_o)(q=0)| &=& \frac{\kappa}{\rho}\frac{2\pi^2 \hbar^2 \rho n_v}{M^*}\nonumber\\
&+& \frac{1}{2}\vert\sum_j e^{2i\phi_{ij'}}[\langle \delta x_{ij'}^2\!+\!\delta y_{ij'}^2\rangle \mathcal{T}_{ij'}+\{\langle \delta x_{ij'}^2\!-\!\delta y_{ij'}^2\rangle \cos 2\phi_{ij'}+ 2\langle \delta x_{ij'} \delta y_{ij'}\rangle \sin 2 \phi_{ij'}\}\mathcal{\tilde{T}}_{ij'}]\mathcal{\tilde{T}}_{ij'} V\vert.\nonumber\\
\end{eqnarray}
\end{widetext}
The thermal correction to $|(\zeta_o + i\alpha_o)(q=0)|$ vanishes in the continuum limit because $\langle \delta x_{ij'}^2 + \delta y_{ij'}^2\rangle$ and $\langle \delta x_{ij'}^2 - \delta y_{ij'}^2\rangle \cos 2\phi_{ij'}+2\langle \delta x_{ij'} \delta y_{ij'}\rangle \sin 2 \phi_{ij'}$ do not depend on $\phi_{ij'}$. Hence Eq.\eqref{EQ:optGapT}.

Among thermal corrections for $\eta_0 (q=0)$, the biggest contribution comes from 
\begin{eqnarray}
&\,&\frac{1}{4}\langle \delta \Delta x_i^2 + \delta \Delta y_i^2\rangle \sum_j\mathcal{T}_{ij'}^2 V\nonumber\\ 
&\approx& \frac{1}{8}\frac{\pi\hbar^2\rho n_v}{M^* \lambda^4}  \langle \delta \Delta x_i^2 + \delta \Delta y_i^2\rangle  \int r dr   K_0 \left(\frac{r}{\lambda}\right)\nonumber\\
&=& \frac{1}{8}\frac{\pi\hbar^2\rho n_v}{M^* \lambda^4} \langle \delta \Delta x_i^2 + \delta \Delta y_i^2\rangle.
\end{eqnarray}
if the optical phonon is nearly gapless. Following Fetter \cite{FETTER1967}, we find the thermal fluctuation due to the gapless optical phonon with dispersion $\omega_o = \eta_o \tilde{d}_o  q/2$ to be
\begin{eqnarray}
\langle \delta \Delta x_i^2 + \delta \Delta y_i^2 \rangle &\approx& \frac{2k_B T}{\pi N_v \hbar \rho}\sum_{\bf q} \frac{\eta_o ({\bf q})}{\omega_o^2 ({\bf q})}\nonumber\\
&=& \frac{8k_B T}{\pi \hbar \rho \tilde{d}_o^2 n_v \eta_o}\int \frac{dq}{q} = \frac{4k_B T}{\pi \hbar \rho \tilde{d}_o^2 n_v \eta_o} \ln N_v.\nonumber\\
\end{eqnarray}
As for other thermal fluctuation terms for $\eta_0 (q=0)$, $\frac{1}{4} \sum_j \langle \delta \Delta x_i \delta \Delta x_j + \delta \Delta y_i \delta \Delta y_j\rangle\mathcal{T}_{ij'}^2 V$ is a positive number of comparable order 
whereas the rest of terms much smaller  ($\sim \ln n_v \lambda^2$). Finally, we note that in calculating the relative coordinate fluctuation, we have only computed optical phonon modes contribution, which is justified as the $\ln N_v$ divergence originate from the infrared divergence of optical phonons. 

\section{Lattice energy summation}

Here, we evaluate Eqs.\eqref{EQ:ESum} and \eqref{EQ:GSum}. We will first compute $\sum_j K_0 (r_{0j'}/\lambda)$ through calculating \cite{FETTER1975} 
\begin{equation}
\sum_j e^{i {\bf k}\cdot {\bf r}_j} K_0 (\lambda^{-1}|{\bf r}_j - {\bm \tau}|).
\end{equation}
Using Fourier transform, we get
\begin{eqnarray}
&\,&\sum_j e^{i {\bf k}\cdot {\bf r}_j} K_0 (\lambda^{-1}|{\bf r}_j - {\bm \tau}|)\nonumber\\ &=& 2\pi n_v \sum_{\bf g} \frac{e^{i({\bf g}+{\bf k})\cdot {\bm \tau}}}{({\bf g}+{\bf k})^2 + \lambda^{-2}}\nonumber\\ 
&=& 2\pi n_v  \sum_{\bf g}\frac{e^{i({\bf g}+{\bf k})\cdot {\bm \tau}}}{({\bf g}+{\bf k})^2}\nonumber\\ 
&-& 2\pi n_v \lambda^{-2}\sum_{\bf g}\frac{e^{i({\bf g}+{\bf k})\cdot {\bm \tau}}}{({\bf g}+{\bf k})^2[({\bf g}+{\bf k})^2 + \lambda^{-2}]},
\end{eqnarray}
where ${\bf g}$ is the reciprocal lattice vector. Note that we have divided up the sum into the first part that is logarithmically divergent at both long- and short-wavelength and the second part that cancels out the long-wavelength divergence and is convergent at short-wavelength. This second part is of order $1/n_v \lambda^2$ once we exclude the ${\bf g} = 0$ term.

To calculate the first part, we first note
\begin{eqnarray}
\frac{1}{({\bf g}+{\bf k})^2} &=& 2\left(\int_0^{1/\sqrt{4\pi n_v}}+\int^\infty_{1/\sqrt{4\pi n_v}}\right) y dy e^{-y^2 ({\bf g}+{\bf k})^2}\nonumber\\ 
&=&  \frac{e^{-({\bf g}+{\bf k})^2/4\pi n_v}}{({\bf g}+{\bf k})^2} + 2 \int_0^{1/\sqrt{4\pi n_v}} y dy e^{-y^2 ({\bf g}+{\bf k})^2}. \nonumber\\
\end{eqnarray}
Then we use Poisson summation
\begin{equation}
4\pi n_v y^2 \sum_{\bf g} e^{i{\bf g}\cdot{\bm \tau}}e^{-y^2 ({\bf g}+{\bf k})^2} = \sum_j e^{-i {\bf k}\cdot({\bm \tau}-{\bf r}_j)}e^{-({\bm \tau}-{\bf r}_j)^2/4y^2}
\end{equation}
(this originates from $\sum_{\bf g} e^{i{\bf g}\cdot {\bf r}'} = \sum_j \delta({\bf r}'-{\bf r}_j)$) to derive
\begin{eqnarray}
&\,&\int_0^{\frac{1}{\sqrt{4\pi n_v}}} y dy \sum_{\bf g} e^{i({\bf g}+{\bf k})\cdot {\bm \tau}}  e^{-y^2 ({\bf g}+{\bf k})^2}\nonumber\\ 
&=& \frac{1}{4\pi n_v} \int_0^{\frac{1}{\sqrt{4\pi n_v}}} \frac{dy}{y}\sum_j e^{i{\bf k}\cdot{\bf r}_j}e^{-({\bm \tau}-{\bf r}_j)^2/4y^2}\nonumber\\
&=& \frac{1}{4\pi n_v}\sum_j e^{i{\bf k}\cdot{\bf r}_j}\int^\infty_{\pi n_v ({\bf r}_j - {\bm \tau})^2}\frac{dt}{t}e^{-t}\left(-\frac{1}{2}\right)\nonumber\\ 
&=& \frac{1}{8\pi n_v} \sum_j e^{i{\bf k}\cdot{\bf r}_j} E_1 [\pi n_v ({\bf r}_j - {\bm \tau})^2].
\end{eqnarray}

Now we can evaluate $\sum_j K_0 (r_{0j'}/\lambda)$. The important step is to isolate out ${\bf g} =0$ part:
\begin{widetext}
\begin{eqnarray}
&\,&\sum_j e^{i {\bf k}\cdot {\bf r}_j} K_0 (\lambda^{-1}|{\bf r}_j - {\bm \tau}|)\nonumber\\
&=& 2\pi n_v  \sum_{\bf g}\frac{e^{i({\bf g}+{\bf k})\cdot {\bm \tau}}}{({\bf g}+{\bf k})^2}- 2\pi n_v \lambda^{-2}\sum_{\bf g}\frac{e^{i({\bf g}+{\bf k})\cdot {\bm \tau}}}{({\bf g}+{\bf k})^2[({\bf g}+{\bf k})^2 + \lambda^{-2}]}\nonumber\\
&=&2\pi n_v \sum_{\bf g} \frac{e^{i({\bf g}+{\bf k})\cdot {\bm \tau}}e^{-({\bf g}+{\bf k})^2/4\pi n_v}}{({\bf g}+{\bf k})^2}+ \frac{1}{2}\sum_j e^{i{\bf k}\cdot{\bf r}_j} E_1 [\pi n_v ({\bf r}_j - {\bm \tau})^2]- 2\pi n_v \lambda^{-2}\sum_{\bf g}\frac{e^{i({\bf g}+{\bf k})\cdot {\bm \tau}}}{({\bf g}+{\bf k})^2[({\bf g}+{\bf k})^2 + \lambda^{-2}]}\nonumber\\
&=& \frac{2\pi n_v e^{i{\bf k}\cdot{\bm \tau}}}{k^2}\left[e^{-k^2/4\pi n_v}-\frac{\lambda^{-2}}{k^2 + \lambda^{-2}}\right] + 2\pi n_v \sum_{{\bf g}\neq 0}  \frac{e^{i({\bf g}+{\bf k})\cdot {\bm \tau}}e^{-({\bf g}+{\bf k})^2/4\pi n_v}}{({\bf g}+{\bf k})^2}+\frac{1}{2}\sum_j e^{i{\bf k}\cdot{\bf r}_j} E_1 [\pi n_v ({\bf r}_j - {\bm \tau})^2],
\end{eqnarray}
\end{widetext}
where we dropped terms of order $1/n_v \lambda^2$ terms in the last step. Taking ${\bf k} \to 0$, we obtain
\begin{align}
\sum_j K_0 (\lambda^{-1}|{\bf r}_j - {\bm \tau}|) = &\sum_j K_0\left(\frac{r_{0j'}}{\lambda}\right)\nonumber\\
= &2\pi n_v \lambda^2 - \frac{1}{2} + \frac{1}{2} \gamma_{HQV}.
\end{align}
Combining this with Fetter's result \cite{FETTER1966a}
\begin{equation}
\sum_{j \neq 0} K_0\left(\frac{r_{0j}}{\lambda}\right) = \pi n_v \lambda^2 - \frac{1}{2} \ln 4\pi n_v \lambda^2 - \frac{1}{2} (1-\gamma_E) + \frac{1}{2}\gamma_{FQV}
\end{equation}
gives us
\begin{align}
\Sigma_0 = & 4 \pi n_v \lambda^2 - \frac{1}{2} \ln 4\pi n_v \lambda^2 - \frac{1}{2} (2-\gamma_E)\nonumber\\
+&\frac{1}{2}\gamma_{FQV} + \frac{1}{2} \gamma_{HQV}.
\end{align}
We also find
\begin{equation}
\Sigma_1 = 4n_v\lambda^2 \frac{d}{d(n_v\lambda^2)}\Sigma_0 = 16\pi n_v \lambda^2 - 2,
\end{equation}
using $dK_0(x)/dx = x[K_0(x)-K_2(x)]/2$.

For the spin current part of the energy, we note
\begin{equation}
\sum_j e^{i{\bf k}\cdot {\bf r}_j} \log(|{\bf r}_j - {\bm \tau}|) = -2\pi n_v \sum_{\bf g} \frac{e^{i({\bf g}+{\bf k})\cdot {\bm \tau}}}{({\bf g}+{\bf k})^2},
\end{equation}
and obtain
\begin{equation}
\Sigma_{sp} = -\frac{1}{2} \log \pi n_v \xi^2 - \frac{1}{2} \gamma_E + \frac{1}{2} \gamma_{FQV} - \frac{1}{2}\gamma_{HQV}.
\end{equation}

\end{document}